\documentclass[acmlarge, authorversion, nonacm]{acmart}

\usepackage{gensymb}
\usepackage{amsmath}

\usepackage{subfig}
\usepackage{tabularx}
\usepackage{booktabs}
\usepackage{float}

\usepackage{balance} 

\AtBeginDocument{%
  }

\setcopyright{acmlicensed}
\acmJournal{IMWUT}
\acmYear{2024} 
\acmVolume{X} 
\acmNumber{X} 
\acmArticle{X} 
\acmDOI{10.XXXX/XXXXXXX}

\begin{document}

\newcommand{\name}{WristSonic}{}

\newcommand{\add}[1]{\textcolor{red}{{#1}}}
\newcommand{\delete}[1]{{\sout{ #1}}}


\title{\name{}: Enabling Fine-grained Hand-Face Interactions on Smartwatches Using Active Acoustic Sensing}


\author{Saif Mahmud}
\affiliation{%
  \institution{Cornell University}
  \city{Ithaca, NY}
  \country{USA}}
\email{sm2446@cornell.edu}
\orcid{0000-0002-5283-0765}

\author{Kian Mahmoodi}
\affiliation{%
  \institution{Cornell University}
  \city{Ithaca, NY}
  \country{USA}}
\email{km777@cornell.edu}
\orcid{0009-0009-7660-1922}

\author{Chi-Jung Lee}
\affiliation{%
  \institution{Cornell University}
  \city{Ithaca, NY}
  \country{USA}}
\email{cl2358@cornell.edu}
\orcid{0000-0002-1887-4000}


\author{Francois Guimbretiere}
\affiliation{%
  \institution{Cornell University}
  \city{Ithaca, NY}
  \country{USA}}
\email{francois@cs.cornell.edu}
\orcid{0000-0002-5510-6799}

\author{Cheng Zhang}
\affiliation{%
  \institution{Cornell University}
  \city{Ithaca, NY}
  \country{USA}}
\email{chengzhang@cornell.edu}
\orcid{0000-0002-5079-5927}

\renewcommand{\shortauthors}{Saif Mahmud et al.}

\begin{abstract}

\begin{figure}[H]
    \centering
    \includegraphics[width=\textwidth]{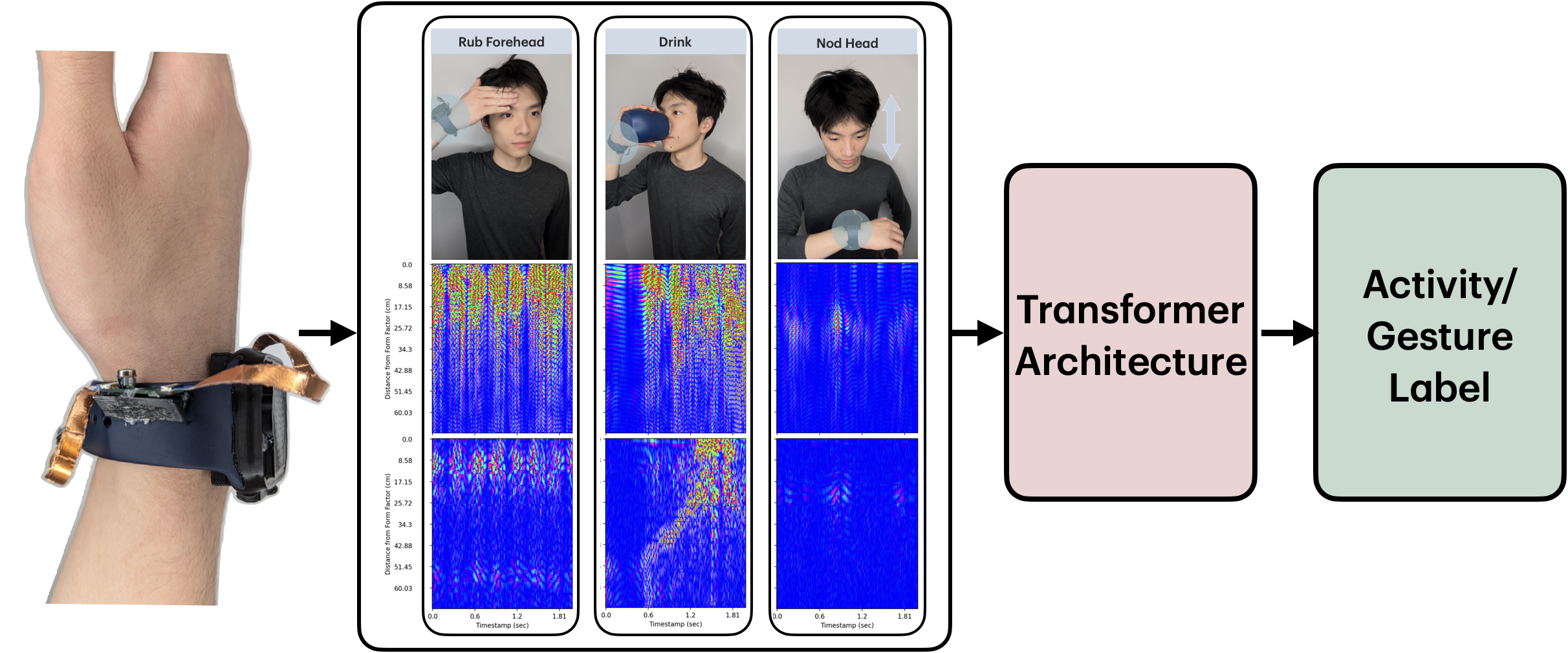}
    \caption{\name{} detects user activities and gestures involving hand-face interaction using an active acoustic sensing system mounted on the user’s wrist. Echo profiles are generated from the acoustic reflection data and fed into a transformer-based model, which classifies the activity being performed (e.g., rubbing forehead, drinking, nodding head).}
    \label{fig:teaser}
\end{figure}


Hand-face interactions play a key role in many everyday tasks, providing insights into user habits, behaviors, intentions, and expressions. However, existing wearable sensing systems often struggle to track these interactions in daily settings due to their reliance on multiple sensors or privacy-sensitive, vision-based approaches. To address these challenges, we propose \name{}, a wrist-worn active acoustic sensing system that uses speakers and microphones to capture ultrasonic reflections from hand, arm, and face movements, enabling fine-grained detection of hand-face interactions with minimal intrusion. By transmitting and analyzing ultrasonic waves, \name{} can distinguish a wide range of gestures—such as tapping the temple, brushing teeth, and nodding—through a Transformer-based neural network architecture. This approach achieves robust recognition of 21 distinct actions with a single, low-power, privacy-conscious wearable. Through two user studies with 15 participants in controlled and semi-in-the-wild settings, \name{} demonstrates high efficacy, achieving Macro F1-scores of 93.08\% and 82.65\%, respectively.
\end{abstract}




\maketitle

\section{Introduction} 
\label{sec:intro}


Hand-face interactions are essential for a wide range of everyday tasks, from adjusting glasses and swiping the chin to scratching an itch on the face and covering the mouth while coughing. These interactions provide critical insights into the user’s habits, behaviors, intentions, and expressions. Therefore, the ability to continuously track these hand-face interactions can significantly advance our understanding of human activities and enhance the interaction experience on wearable devices.

However, tracking these interactions with existing wearable sensing systems is challenging. Hand-face interactions typically involve complex postures and movements of the hands, arms, and corresponding face deformations. To accurately recognize a wide range of detailed hand-face interactions, a sensing system must capture information from these body parts simultaneously, which poses a significant challenge for wearables. Wrist-mounted devices generally only track hand or arm movement, while head-mounted wearables primarily track facial or head movements. Achieving reliable performance often requires multiple sensors on both the head and arm to recognize hand-face touch behaviors~\cite{voice-accompanied}. Additionally, many of these systems rely on egocentric cameras~\cite{FaceSight}, which are energy-intensive and raise privacy concerns for wearable applications. Therefore, there is a clear need for a single, minimally obtrusive, low-power, and privacy-aware wearable device capable of recognizing a wide range of hand-face interactions involving the arms, hands, and face.

Given these limitations, we ask a straightforward research question: \textit{Can we enable fine-grained hand-face interaction by simultaneously tracking subtle hand movements and the corresponding face deformations that occur during hand-face interactions?} An affirmative answer to this question would provide a significant advantage, enabling smartwatches to facilitate seamless interaction through various gestures and activities. Furthermore, if such a solution operates with low power consumption and maintains privacy, it could potentially integrate smoothly with existing systems and drive wider adoption.

In this paper, we introduce \name{}, a wristband-based solution that tracks a broad range of fine-grained hand-face interactions by monitoring arm, head, and face movements through active acoustic sensing. Equipped with speakers and microphones, various wearable devices have been shown capable of tracking body movements via active acoustic sensing~\cite{EchoWrist, mahmud2024actsonic, eario, echospeech, c-fmcw, SonicASL}. Leveraging this capability, \name{} uses ultrasonic waves to model both hand movement and subtle face deformations, allowing for simultaneous tracking of hand-face interactions.

\name{} operates with just two pairs of off-the-shelf speakers and microphones—one facing outward from the hand and the other directed towards the body—to transmit C-FMCW~\cite{c-fmcw} encoded ultrasonic waves and capture their reflections from the arm, head, and face, as the hand interacts with the face. Using a Transformer-based neural network architecture~\cite{ViT}, we process these acoustic reflections to infer various hand-face gestures, daily activities, and head motions. This includes gestures such as touching the earlobe, tapping the temple, drinking, brushing teeth, nodding, and shaking the head, among 21 distinct actions. To evaluate the efficacy of \name{}, we conducted two user studies with 15 participants: one in a controlled lab setting with 10 subjects and another in a naturalistic semi-in-the-wild environment with 5 participants. \name{} demonstrates robust performance across different conditions, achieving macro F1-scores of 93.08\% and 82.65\% in lab and semi-in-the-wild settings, respectively.

Our contributions in this work are summarized as follows: 
\begin{itemize} 
    \item We present the first single smartwatch-based system leveraging active acoustic sensing to distinguish a wide range of fine-grained hand-face interactions by simultaneously sensing information from the hand and face.
    \item We implemented a self-attention-based architecture to model the received acoustic reflections, enabling the recognition of 21 distinct hand-face interactions. 
    \item We conducted two user studies to evaluate the system’s performance across various scenarios and environmental conditions, demonstrating its performance, robustness, and usability. 
\end{itemize}
\section{Related Work} 
\label{sec:rel-works}

Many systems have been developed to track hand-face interaction or touching behavior across different wearable devices. This section examines current approaches to tracking hand-face interactions, first exploring solutions on non-smartwatch form factors, such as eyeglasses, earbuds, and neckbands, and then examining solutions proposed specifically for smartwatches.

\subsection{Form Factors Beyond Smartwatches}

Touching or interacting with the face is a natural, frequent behavior, often occurring unconsciously or habitually, and sometimes related to behavioral concerns like fidgeting or nail-biting~\cite{tanaka2008nailbiting}. Additionally, excessive face-touching, especially in the T-zone~\cite{lucas2020frequency, rahman2020frequently}, has raised health concerns, particularly in light of pandemic-related hygiene.

Frontal cameras on mobile devices~\cite{SideSight, imaginary-phone-Holz-Baudisch, imaginary-interfaces-Patrick-Baudisch, HandSee} are frequently used to track hand-face interactions through computer vision algorithms~\cite{sun2014deep, parkhi2015deep, deng2019arcface}. Beyond mobile devices, IR cameras or optical sensors on wearables like eyeglasses~\cite{FaceSight, CheekInput} and neck-mounted devices~\cite{D-Touch} have also been employed to detect hand-face behaviors. While these vision-based methods perform well in controlled settings for activities such as tracking face-touch positions or actions like drinking and eating, they often require substantial computational power, which limits their use in an everyday setting. Additionally, challenges such as hand-over-face occlusion and variable lighting reduce their robustness in real-world scenarios. Radar-based methods, such as those using mmWave radar~\cite{Soli, Soli-interact}, offer another approach, allowing gesture recognition without light dependency but facing similar occlusion challenges and high power demands.

Acoustic-based systems on glasses~\cite{mahmud2024actsonic, MunchSonic, echospeech} or earbuds~\cite{MAF, eario, EarBuddy, face-touch-sonar, FaceOri} have used face deformation as a proxy for detecting hand-face interactions. These systems can infer activities involving both face and hand, such as eating, drinking, and brushing teeth, or gestures near or on the face, like pinching or scratching, by detecting deformations. However, relying on visible face deformation to track hand interactions can introduce usability issues, including potential social discomfort and user fatigue~\cite{hand-face-design-space}. Moreover, subtle, fine-grained hand-face interactions may not cause detectable deformations that these systems can track effectively.

In addition to vision- and acoustic-based sensing, there are thermal and physiological sensing methods~\cite{FaceSense} on ear-worn devices that track facial touch zones. Electrooculography (EOG)-based sensing on eyeglasses~\cite{Itchy-nose} has also been used to detect specific nose movements, such as flicking or rubbing.

However, none of these approaches on various wearables can simultaneously track both hand and face movements, which is essential for the precise detection of hand-face interactions. These systems typically rely on indirect information, either from face deformations or from detecting hand movements toward the face, making it challenging to capture fine-grained hand-face interactions that require tracking both hand and face in tandem.

\subsection{Smartwatch-Based Solutions} 

Smartwatches have become versatile wearables that support a wide range of hand-face interactions through various sensing modalities. Using IMUs~\cite{smartwatch-arm-tracking, voice-accompanied} or low-power capacitive sensors~\cite{socially-acceptable-hand-face, CapBand}, smartwatches can track gestures such as pinching the ear rim, touching the vocal cords, or making a “shushing” gesture. However, these IMU- or capacitance-based systems cannot capture facial deformations alongside gesture tracking, which limits their effectiveness in fine-grained tracking.

To address smartwatch usability challenges from limited screen space, gesture recognition approaches have extended the input area to nearby skin surfaces using camera or RF-based sensing~\cite{Skinput, SkinTrack, skin-buttons, AuraSense, TapSkin, no-handed-smartwatch-interaction}. However, these systems typically focus on wrist-adjacent skin surfaces and do not support broader hand-face interactions.

Head-based interactions, such as gaze tracking~\cite{Orbits, gaze-interact}, have been explored to enable hands-free smartwatch control. More recently, mmWave and IMU-based techniques have enabled head gestures like nodding or shaking~\cite{Headar, Nod-to-Auth} to serve as interactive input methods. Additionally, acoustic sensing, which leverages the growing presence of speakers and microphones in smartwatches, has extended the interaction capabilities of these devices. Ultrasonic reflection-based approaches~\cite{EchoWrist, fingerio, FingerPing, EchoFlex} and Doppler-based methods~\cite{SoundWave, SonicOperator} have demonstrated robust hand movement detection by using acoustic reflections on the skin. Device-free acoustic systems~\cite{Strata, AudioGest, CAT} have also shown effective tracking of detailed hand movements in close proximity. During the COVID pandemic, an NFC-based smartwatch system~\cite{NoFaceContact} was proposed to alert users when they attempt to touch their faces, aiming to reduce viral spread.

Despite these innovations, a primary limitation of current systems—both smartwatch- and non-smartwatch-based—is their focus on tracking a single component, such as either face deformation or isolated hand movements. This limits their ability to capture the full complexity of hand-face interactions, which requires tracking both hand and face dynamics for more precise interaction detection. Smartwatches, equipped with active acoustic sensing, present a unique opportunity to track both hand and face movements simultaneously, especially when the hand is near the face. \name{} leverages this dual-tracking capability, introducing a new paradigm of interaction that incorporates hand-to-face gestures, activities, and head motion tracking on smartwatches.
\section{Implementation of \name{} Sensing System}
\label{sec:implementation}

C-FMCW~\cite{c-fmcw}-based active acoustic sensing has been proven very efficient in tracking different body parts through its utilization in various wearable systems such as eyeglasses~\cite{mahmud2024actsonic, PoseSonic, echospeech}, earbuds~\cite{eario, SonicASL}, and smartwatches~\cite{EchoWrist}. These systems transmit ultrasonic waves and capture their reflections to model nearby body parts and objects. Given their high spatial and temporal resolution and low power consumption, we adopted a similar sensing technique tailored specifically for the hand-face interaction task on a smartwatch in \name{}.

\subsection{Active Acoustic Sensing Preliminaries}
\name{} uses Cross-correlation-based Frequency Modulated Continuous Wave (C-FMCW)~\cite{c-fmcw} ultrasonic sensing to detect motion and infer activities. Its two transmitters emit chirps across frequency ranges of 18.0–21.0~KHz and 21.5–24.5~KHz. These signals reflect off the user’s body, and receivers capture the reflected signals, which are time-shifted due to the motion. The received signals are filtered and cross-correlated with the transmitted signals, generating an echo frame that represents areas of activity—higher values in the echo frame correspond to more significant motion.

To isolate dynamic movements while filtering out static elements, consecutive echo frames are subtracted, creating differential echo frames. With \name{}'s two transmitters and two receivers, this setup captures four channels of data, yielding four differential echo profiles per captured data point. This differential approach enhances the system’s ability to track real-time hand-face interactions effectively.

\subsection{Design Principle for Simultaneous Tracking of Hand and Face}

\begin{figure}[h!]
    \centering
    \includegraphics[width=0.95\textwidth]{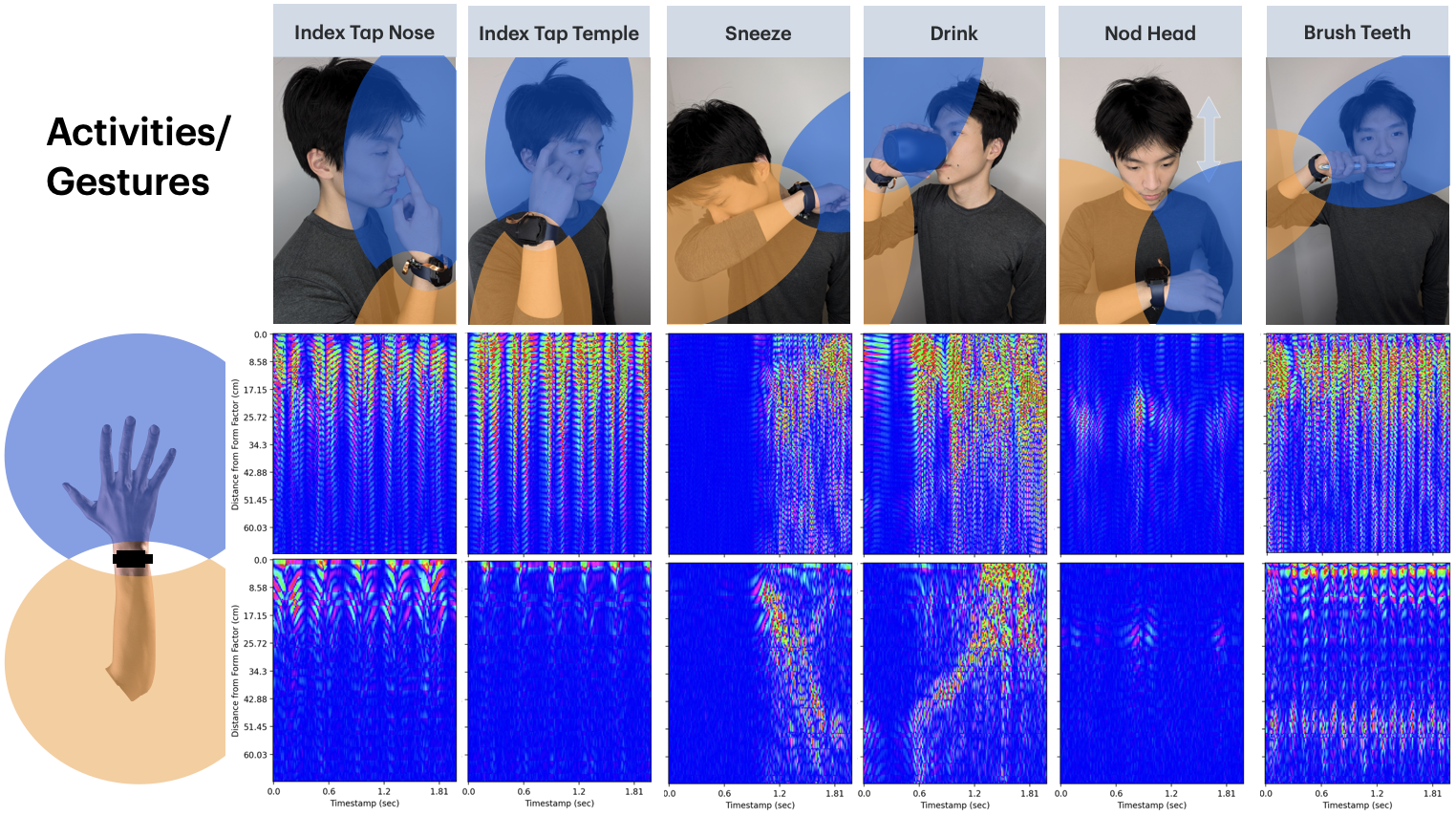}
    \caption{Echo profiles of selected activities over 2.0 seconds of tracking. The top image shows the activity, while the bottom displays the generated echo profile. Blue regions represent areas tracked by the outward-facing sensor (relative to the arm), while orange regions represent those tracked by the inward-facing sensor. The top echo profile corresponds to the outward-facing sensor and the bottom profile to the inward-facing sensor.}
    \label{fig:sensing-principle}
\end{figure}

\subsubsection{Finding an optimal position and orientation for \name{}'s acoustic sensors}
\label{sec:how-optimal-pos}

A key factor in accurately modeling fine-grained hand-face interactions is the simultaneous tracking of both hand and face—a challenge that existing systems struggle to address. To enable \name{} to achieve this, we optimized sensor placement to accommodate the dynamic, ego-centric nature of wrist movements, ensuring robust signal capture despite increased motion and body positioning variability. We conducted a series of pilot studies to determine the optimal position and angle of the sensors relative to the user's arm.

In the initial position-finding study, researchers tested various sensor configurations: both sensors transmitting in the same direction (either outward from the hand or toward the body), and sensors positioned to face opposite directions. For each configuration, ten minutes of data were collected while researchers performed a set of gestures and activities. Differential echo profiles were visualized, and the configuration that produced the most distinct, repeatable, and information-rich echo profiles was identified, using a lightweight classifier trained on this data. 

A similar approach was used to finalize the sensor orientation angle. Data was collected with the sensors positioned at both 45\textdegree~and 90\textdegree~angles relative to the arm. Following the same evaluation process, the 90\textdegree~tilt provided the most consistent and informative echo profiles. Consequently, the final sensing configuration uses two opposite-facing transceivers: one facing outward from the hand (illustrated in blue in Figure~\ref{fig:sensing-principle}) and the other directed toward the user’s body (indicated in orange in Figure~\ref{fig:sensing-principle}).

\subsubsection{How does the sensing placement enable simultaneous tracking of hand movement and face deformation?}

The core design principle behind \name{}'s sensor placement is to enable simultaneous tracking of hand movement and face deformation during hand-face interactions. The finalized position and orientation, as detailed in Subsec.~\ref{sec:how-optimal-pos}, allows the sensor facing the body (orange in Figure~\ref{fig:sensing-principle}) to capture arm movement, indicating the approach of the arm towards the face. Meanwhile, the outward-facing sensor (blue in Figure~\ref{fig:sensing-principle}) captures subtle finger movements and any accompanying face deformation.

Analysis of motion patterns captured by the acoustic sensors reveals that the outward-facing sensor (blue) picks up stronger reflections due to its proximity to the face, capturing subtle hand movements in close proximity. The body-facing sensor (orange) focuses on tracking arm movement, providing insight into how the arm moves toward the face and its activity during various gestures and interactions.

For instance, in the act of drinking, the outward-facing sensor captures face movements associated with drinking, while the body-facing sensor tracks the arm's position as it holds the cup. A similar pattern is observed in activities like brushing teeth, where repetitive arm movements are captured. Additionally, subtle gestures, such as tapping the nose or temple, generate unique face deformation signals depending on the contact location, while the body-facing sensor registers corresponding arm movements. 

In the case of head motion gestures, performed while the hand is not in direct contact with the face, the user might look at the smartwatch with movements like nodding or shaking. Here, the outward-facing sensor captures the mid-air head motion, while the body-facing sensor detects slight arm movements associated with head motion.

Thus, \name{} is capable of simultaneously tracking both face and hand movements, enabling fine-grained modeling of hand-face interactions using the learning system specifically designed for acoustic sensing data captured by \name{}.

\subsection{\name{} Hardware and Form Factor}

\begin{figure}[h!]
    \centering
    \includegraphics[width=\textwidth]{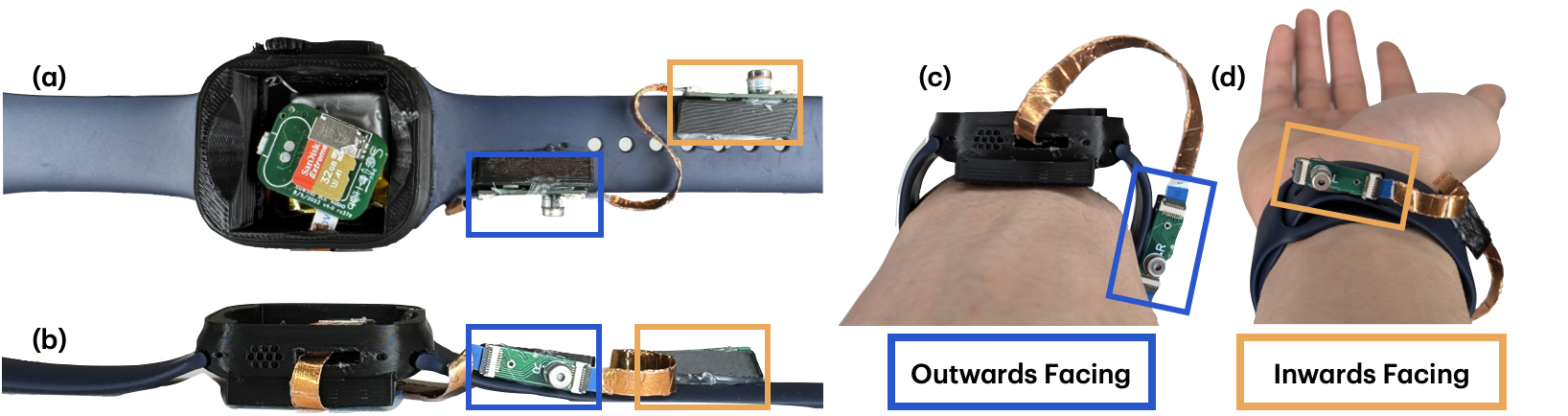}
    \caption{Hardware of \name{}. (a) and (b) show top-down and side views of \name{}'s form factor, with the outward-facing sensor in blue and the inward-facing sensor in orange. (c) Knuckle-up view, highlighting the outward-facing sensor. (d) Palm-up view, highlighting the inward-facing sensor.}
    \label{fig:hardware-formfactor}
\end{figure}

As illustrated in Figure~\ref{fig:hardware-formfactor}, the hardware architecture of \name{} consists of a pair of ultrasonic speakers and microphones. The system includes two identical boards, each equipped with an ICS-43434 microphone to receive signals sent by an OWR-05049T-38D speaker. The controller unit of \name{} is a custom board housing a low-power nRF52840 microcontroller and incorporating two MAX98357A audio amplifiers for ultrasonic FMCW chirp transmission. Captured data is stored on an SD card interfaced with the controller board. Flexible Printed Circuit (FPC) cables connect the sensing boards to the controller unit, and copper shielding on these cables reduces potential interference. The controller unit is powered by a 290mAh LiPo battery, while the entire sensing system draws 96.48 mW of power during data collection, operating at 4.02V and 24.0 mA.

To maintain consistent sensor positioning relative to each user’s arm—a critical factor for reliable data capture—\name{} employs a flexible design to accommodate users with varying wrist sizes. During prototyping, a custom Velcro-based wristband was developed, with sensors mounted on adjustable Velcro strips. This setup allows the wristband to fit a range of users, enabling flexible sensor positioning to ensure consistent placement across users.

For the final prototype, we created a 3D-printed housing shaped like an Apple Watch 6, utilizing an original Apple Watch 6 strap. This housing accommodates the controller unit and battery, while the sensing boards are mounted on the strap, ensuring a secure fit for consistent data recording. This design also highlights the potential for \name{} to integrate seamlessly with commodity smartwatch straps and bodies, requiring minimal modifications.

\section{Deep Learning Framework}
\label{sec:dl-model}

In this section, we present the deep learning framework of \name{}, which processes active acoustic signals to recognize hand-face interaction gestures and activities. Fig.~\ref{fig:ml-model} shows the neural network architecture specifically designed for the \name{} system.

\subsection{Model Architecture}

\begin{figure}[h]
    \centering
    \includegraphics[scale=0.375]{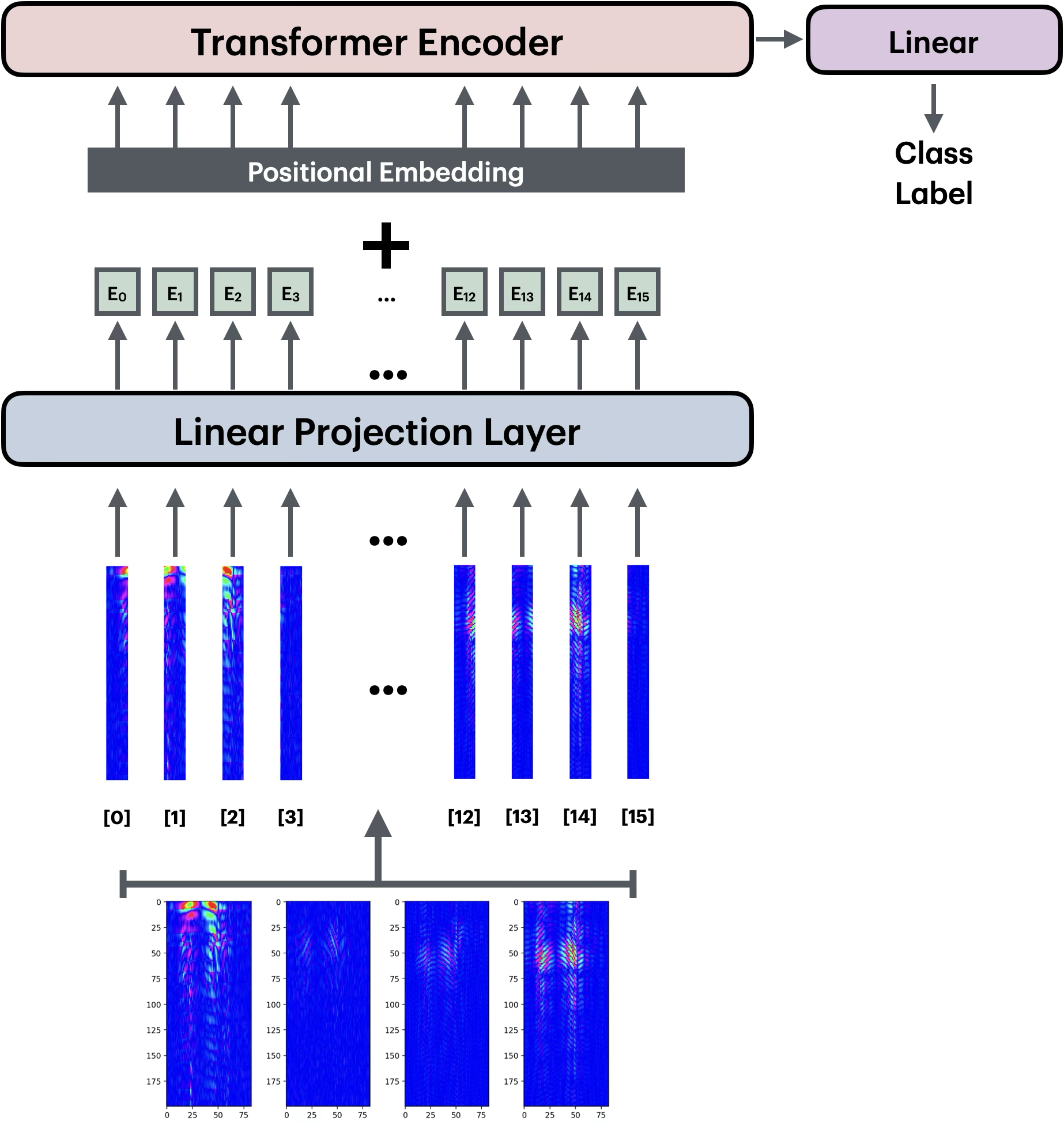}
    \caption{Deep learning model architecture for \name{}.}
    \label{fig:ml-model}
\end{figure}

The design of \name{}'s model architecture draws from recent advances in transformer-based models, particularly the Vision Transformer (ViT)~\cite{ViT} and Audio Spectrogram Transformer (AST)~\cite{AST}. These models use self-attention blocks to capture complex spatiotemporal dependencies in data, which has proven effective in handling sequential inputs such as audio spectrograms and images. Inspired by these techniques, \name{} employs a transformer encoder, which consists of multiple self-attention layers, to model the differential echo profiles.

In our approach, we use sliding windows of the differential echo profile as input to the \name{} model. Through hyperparameter tuning, we found that a sliding window length of 1.00 seconds with a 50\% overlap yields the best performance for gesture and activity recognition tasks. This 1.00-second window corresponds to 83 samples on the temporal axis. Given the structure of our sensing system, the differential echo profile contains data from four channels—one for each reflection path between the speaker-microphone pairs. Therefore, each sliding window input has a shape of $(4 \times 200 \times 83)$, where 200 represents the acoustic range, which spans approximately 68.6 cm from the smartwatch.

To prepare this data for processing in the transformer model, we divide each channel of the differential echo profile into four non-overlapping patches. For each channel, we select a random starting index within the first few samples (in the range $[0, 4]$) and extract the next 79 samples to create a window of $(200 \times 80)$. This windowing step ensures consistency in tensor dimensions while also incorporating a slight temporal shift for better model robustness. Dividing each window into four non-overlapping sections results in 16 patches in total (4 patches per channel). Each patch is then flattened into a 1D vector of size 4000.

After flattening, each patch is mapped to a smaller vector (embedding) of size 768 using a \textit{Linear Projection Layer}, as shown in Fig.~\ref{fig:ml-model}. This projection layer, implemented as a feedforward neural network with Leaky ReLU activation, reduces the high-dimensional patch data into a compact embedding vector, facilitating efficient processing within the transformer. The embedding dimension of 768 is selected based on the design of ViT~\cite{ViT}. Consistent with ViT~\cite{ViT} and BERT~\cite{BERT}, we prepend a \texttt{[CLS]} token to the sequence of patch embeddings, which allows the transformer to generate a single representation for the entire sequence.

To capture positional information within the sequence, we add a positional embedding to each patch embedding. This positional embedding, taken from the standard 1D positional encoding proposed in the original Transformer~\cite{Transformer}, enables the model to maintain the temporal order of patches in the sequence. The positionally encoded embeddings are then processed by the \textit{Transformer Encoder}~\cite{Transformer}, which in this case consists of two attention blocks, with eight attention heads. The output corresponding to the \texttt{[CLS]} token—a vector of size 768—is fed into a final linear layer for classification. Finally, a softmax function is applied to the output logits to yield a probability distribution over possible activities and gestures.

\subsection{Implementation and Training Details}

The \name{} model is trained with focal loss~\cite{lin2017focal}, a variant of cross-entropy loss designed to handle class imbalance. We use the Adam optimizer~\cite{kingma2014adam} with a cosine annealing learning rate scheduler, starting with an initial learning rate of $10^{-2}$. Dropout is applied to the \textit{Linear Projection Layer} with a probability of 0.25 and to the classifier’s linear layer with a probability of 0.20, to prevent overfitting. The model is implemented using PyTorch and PyTorch Lightning and is trained for 50 epochs with a batch size of 64 on GeForce RTX 2080 Ti GPUs.

To evaluate the model’s performance in recognizing gestures and activities, we use the Macro F1-score, a standard metric for multiclass classification tasks. We compute this score using the implementation provided in the TorchMetrics library.
\section{User Study Evaluation Overview}
\label{sec:user-study}

In this section, we describe the user studies conducted to evaluate the performance of \name{}. We began with an elicitation study involving six participants to develop a set of activities and gestures for tracking. Following this, we conducted two user studies with a total of 15 participants: one in a controlled lab environment and the other in a naturalistic, semi-in-the-wild setting. The goal of the in-lab study was to verify the proof-of-concept of \name{} in a controlled experimental setup, while the semi-in-the-wild study provided insights into the feasibility of real-world deployment.

\subsection{Design of Gesture and Activity Set}

\begin{figure}[h]
    \centering
    \includegraphics[scale=0.55]{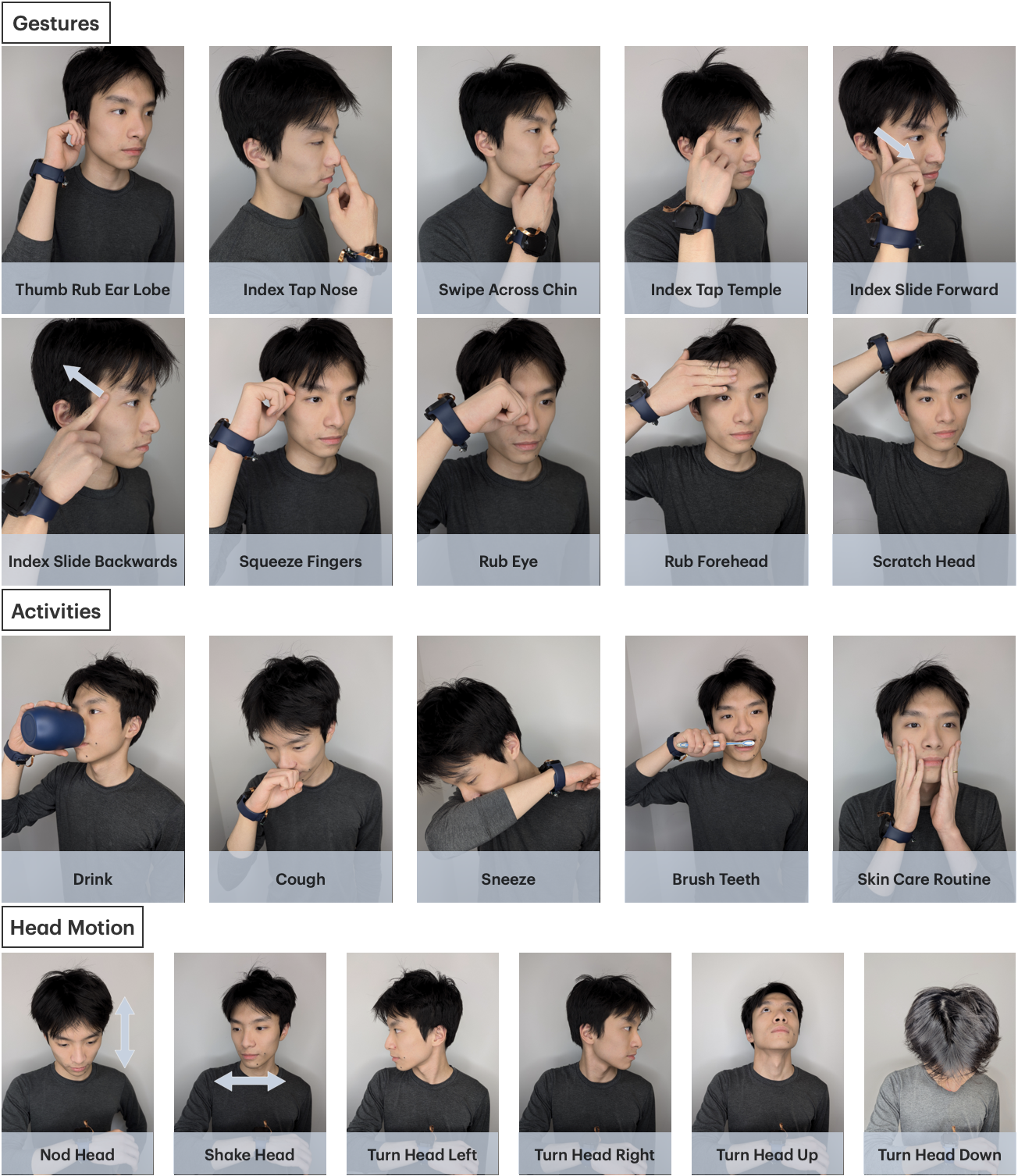}
    \caption{Set of gestures, activities, and head motions tracked by \name{}.}
    \label{fig:act-set}
\end{figure}

Gestures involving the head or face and hand have been widely regarded in prior research as natural, unobtrusive, and socially acceptable forms of wearable interaction~\cite{hand-face-design-space}. Research has shown that it is feasible to implement gesture interactions between the hands and specific regions of the face, such as the ear~\cite{EarTouch}, cheek~\cite{hand-face-design-space, CheekInput}, or nose~\cite{Itchy-nose}. Additionally, prior work~\cite{mahmoud2011interpreting} has found that, for subtle hand-face interactions, gestures targeting the lower face region tend to be more favorable than those involving the upper face.

Building on these design principles from prior studies, we conducted an elicitation study with six participants to guide the creation of a tailored set of activities and gestures to be tracked by the \name{} system. The study was carried out in a controlled lab environment, where we could observe participant interactions closely. Based on findings from earlier research, we initially designed a collection of 28 candidate activities and gestures that might be suitable for tracking through the \name{} system.

Using four core design criteria—usability, social acceptance, fatigue, and the system’s ability to distinguish among gestures (disambiguation)—we refined this preliminary set into a smaller group. We ultimately selected 10 hand-face gestures, 6 head motions, and 5 combined hand-face activities for tracking with \name{}. To ensure the suitability of each choice, we gathered feedback from participants on the practicality and comfort of each gesture, as well as evaluations from researchers on its potential social acceptability and ease of recognition by the sensing system.

After carefully analyzing this feedback and researcher evaluations, we finalized a set of gestures and activities that best met the design criteria. Additionally, we included a "null" label to account for any activities or gestures outside of those defined in the final set. The complete set of gestures and activities, along with this labeling system, is illustrated in Figure~\ref{fig:act-set}.

\subsection{Data Collection Apparatus and Participants}
The \name{} user study was approved by our organization’s Institutional Review Board for Human Participant Research (IRB). We recruited 15 participants, 9 identifying as male and 6 as female, with an average age of $22.667 \pm 3.579$~years (range: 19 to 29 years). The participants had an average height of $169.545 \pm 7.498$~cm and an average weight of $64.994 \pm 12.935$~kg. All the participants in the study reported the right hand as their dominant hand. Upon completing the study, each participant received a \$15 USD gift card.

Sensor data was collected using the \name{} sensing system, mounted on a smartwatch form factor. Participants sat at a table, with video recordings captured via a MacBook Pro camera, which also provided visual instructions. During locomotion activities (i.e., walking), participants received audio instructions through Apple AirPods Pro. For activities such as brushing teeth and drinking, participants were provided with a new Oral-B toothbrush and a cup for the controlled in-lab study, while in the semi-in-the-wild study, participants used their own bottles for drinking. Other activities and gestures required no additional items.

We evaluate the performance of \name{} using the data collected in the user studies. We adopt a leave-one-participant-out evaluation protocol to evaluate performance and ensure a user-independent evaluation. We first discuss the performance of \name{} in a lab setting and subsequently, report the results of the semi-in-the-wild user study in a naturalistic setting.

\section{User Study - 01: In-lab Study}
\subsection{Study Procedure}
We conducted an in-lab user study with 10 participants, each lasting about one hour. At the beginning of the study, we introduced the participants to the gestures and activities through instructional videos and visual demonstrations, allowing them to become fully familiar with each gesture and activity before starting. The study was structured into three main sections: hand-face gestures, general activities, and head motions.

For each section, we collected data over five sessions. In each session, participants were asked to perform a series of gestures in a random order from the recognition set. Each gesture or activity was repeated three times within a session, totaling 15 repetitions per gesture by the end of the study. Between each session, the \name{} form factor was remounted by having participants remove and re-wear the wristband, ensuring variability in placement for more robust data collection.

Participants were seated in front of a camera throughout the study. During the activities and gestures, a researcher remained in the room to monitor the process. However, participants received only minimal guidance on how to perform each gesture, allowing them to execute the gestures in a natural way. This in-lab study produced a total of 48 minutes of interaction data for each participant.

\subsection{Results}
\begin{figure}[h]
    \centering
    \includegraphics[scale=0.5]{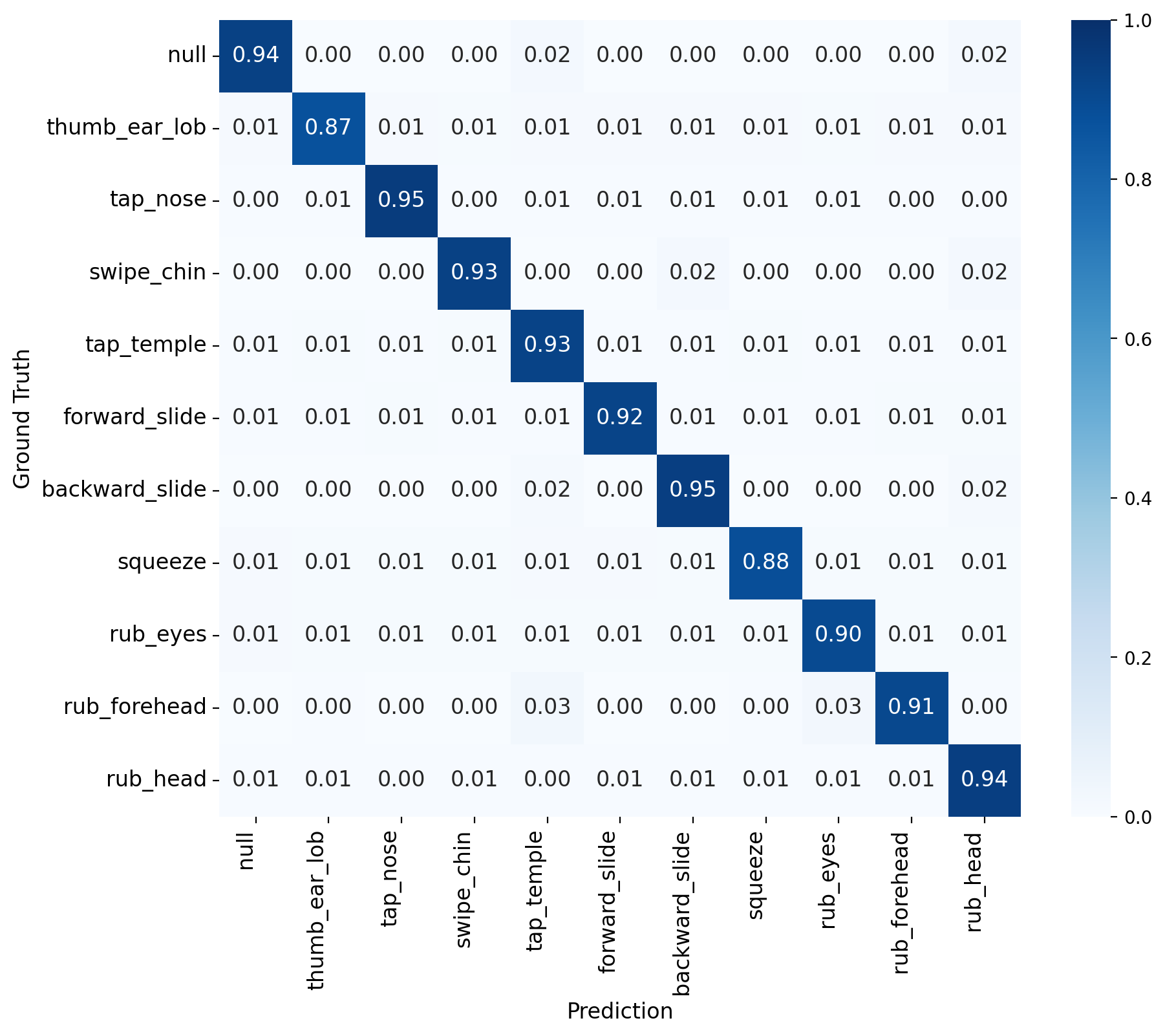}
    \caption{Normalized confusion matrix of gestures in leave-one-participant-out user evaluation of the \textbf{in-lab study}.}
    \label{fig:conf-mat-gest-lab}
\end{figure}

\begin{figure}[h]
    \centering
    \includegraphics[scale=0.55]{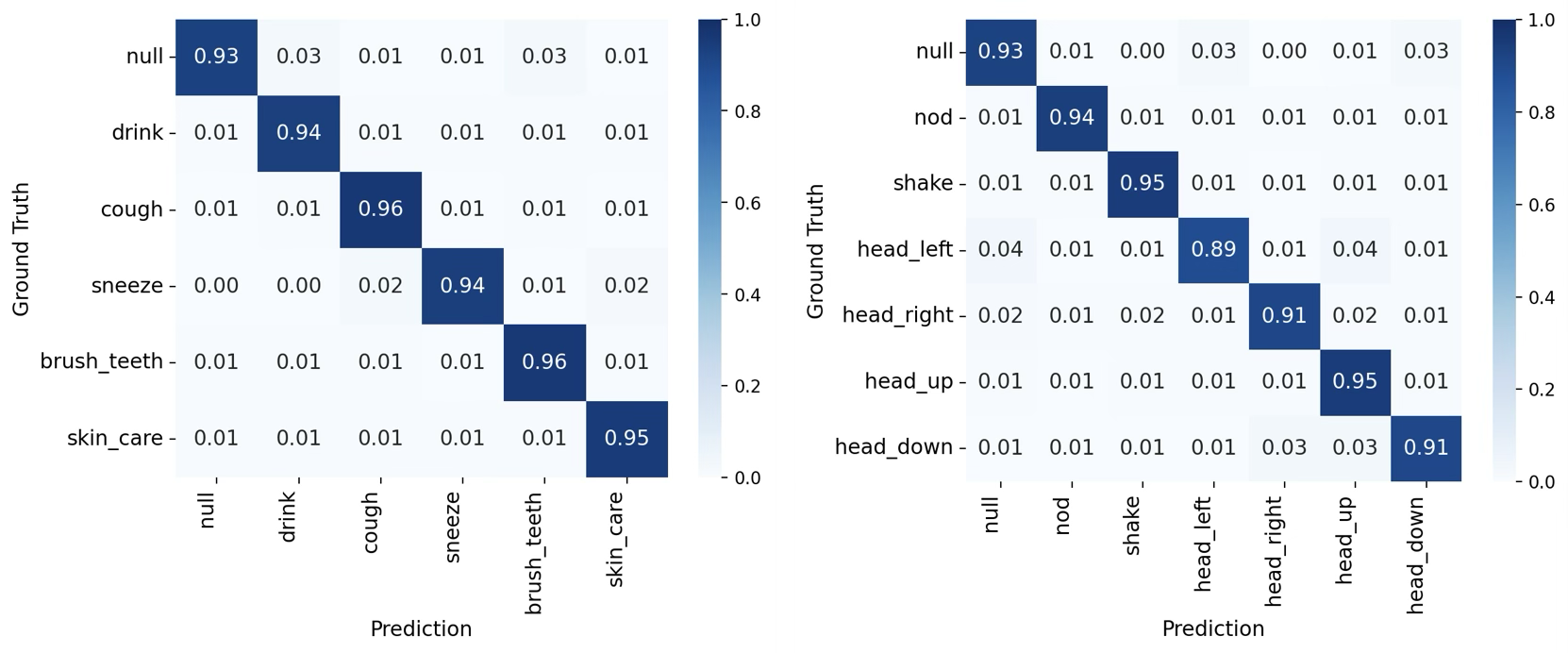}
    \caption{Normalized confusion matrices of everyday activities (left) and head motions (right) in leave-one-participant-out user evaluation of the \textbf{in-lab study}.}
    \label{fig:conf-mat-act-hm-lab}
\end{figure}

We conducted a leave-one-participant-out evaluation on data collected from 10 participants in an in-lab study, where each participant contributed 15 sessions of data. These sessions were divided into three categories: 5 for gestures, 5 for activities, and 5 for head motions. In this evaluation protocol, we trained a model on all sessions of data from all participants except the one in the test set. This approach ensures the model does not encounter data from the test participant, allowing for user-independent evaluation.

We report the averaged normalized confusion matrices across all test participants in Figure~\ref{fig:conf-mat-gest-lab}~and~\ref{fig:conf-mat-act-hm-lab}. The mean macro F1-scores across all participants were 92.14\%, 94.65\%, and 92.45\% for hand-face gestures, everyday activities, and head motions, respectively, with an overall aggregated mean macro F1-score of 93.08\% for the in-lab study.

Examining the confusion matrices in Figure~\ref{fig:conf-mat-gest-lab}~and~\ref{fig:conf-mat-act-hm-lab} reveals that only the "squeezing" gesture and "head rotation to the left" motion yielded slightly lower than 90\% accuracy. Analysis of sensor signals fed into the learning pipeline showed that the squeeze motion involved subtle finger movements without significant face deformation, leading the model to occasionally confuse it with other activities, resulting in a few random inferences. Additionally, all users wore the \name{} form factor on their right wrist, and head movements were performed while looking at the \name{} device. Observations from the study data suggest that left head rotation caused slight movement in other body parts compared to other head movements, particularly in the right hand, which is tracked by the sensing system. This slight motion interference led to a marginally reduced accuracy for left head rotations compared to other head motions.

\section{User Study - 02: Semi-in-the-wild Study}
\subsection{Study Protocol}
We conducted a semi-in-the-wild user study with 5 participants to assess the efficacy of \name{} in realistic scenarios. None of the participants in this semi-in-the-wild study had taken part in the first in-lab study. This semi-in-the-wild study was divided into three sections, each representing typical real-world user movement and environmental noise conditions. These sections were as follows:

\begin{itemize}
    \item \textbf{Indoor user movement:} Since \name{} relies on acoustic reflections to model user movements, it is naturally impacted by locomotion, such as walking. To evaluate the robustness of \name{} in this context, participants were asked to walk randomly while performing gestures and activities in a hallway. Unlike in the controlled lab environment, they did not have access to video instructions. Instead, they received audio instructions via Apple AirPods Pro, allowing them to focus on movement without interruption.
    \item \textbf{Exposure to café or restaurant noise}: To assess \name{}’s resilience to ambient noise and surrounding people’s movements, data was collected while participants sat alone at a café or restaurant table. Although they did not eat, participants drank water as part of the study to simulate some degree of real interaction.
    \item \textbf{Exposure to roadside noise while walking curbside:} This section evaluated the effect of outdoor noise and walking on \name{}’s performance. Participants received instructions in the same manner as in the indoor walking scenario—through audio prompts via AirPods. Although performing activities like brushing teeth or skincare by the curbside may seem unusual, we included these actions for consistency across environments.
\end{itemize}

We collected data in three sessions for each section, with participants repeating each activity or gesture five times per session. Additionally, we followed the same remounting protocol used in the in-lab study, where participants removed and re-wored the wristband between sessions to introduce minor variability in placement.

\subsection{Results}

\begin{figure}[h]
    \centering
    \includegraphics[scale=0.5]{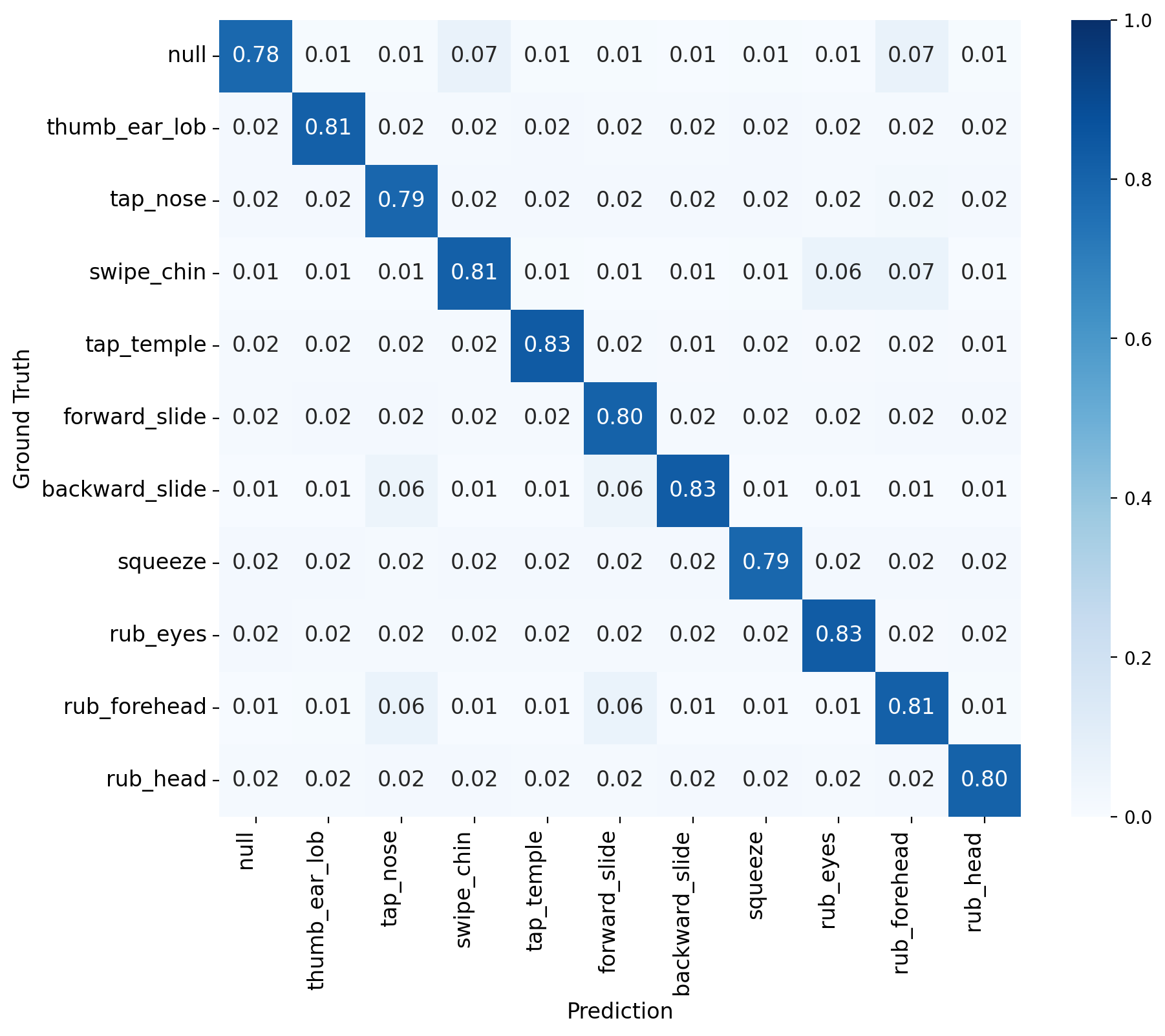}
    \caption{Normalized confusion matrix of gestures in leave-one-participant-out user evaluation of \textbf{semi-in-the-wild study}.}
    \label{fig:conf-mat-gest-wild}
\end{figure}

\begin{figure}[h]
    \centering
    \includegraphics[scale=0.55]{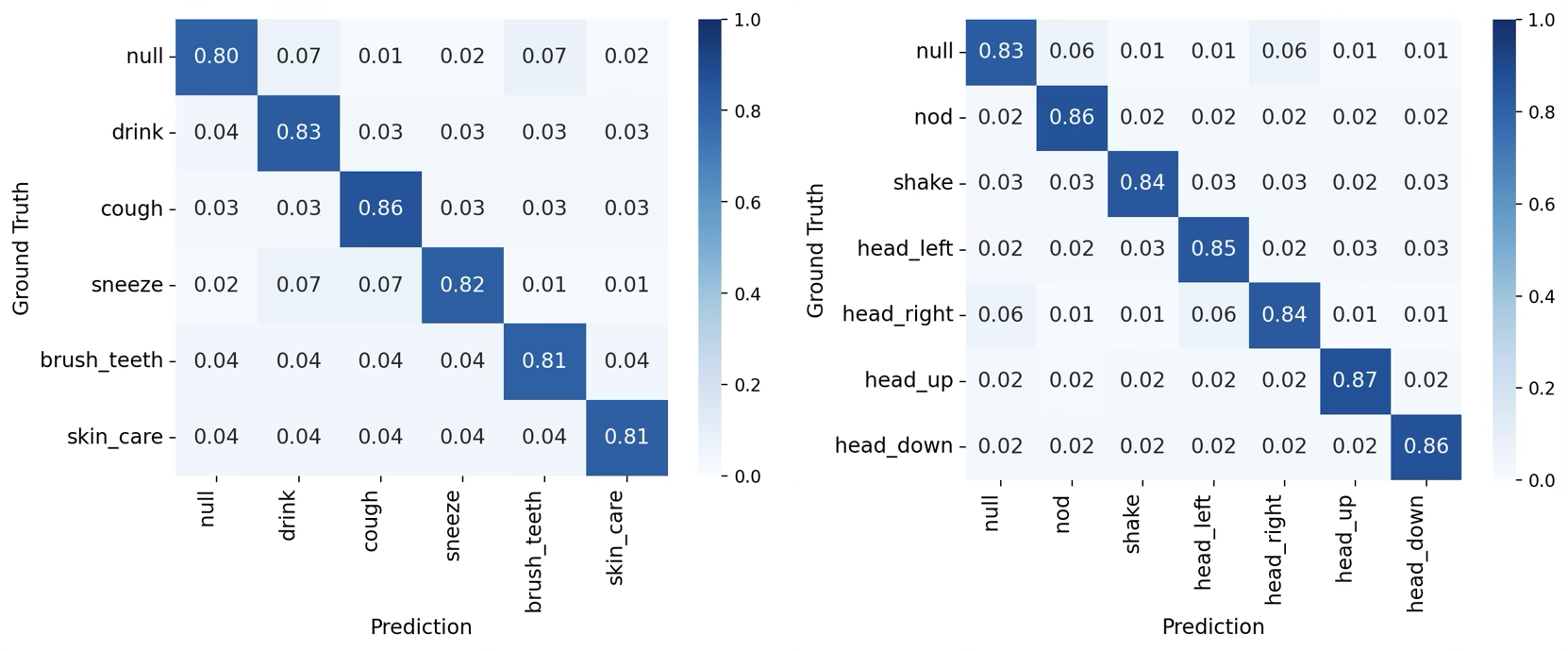}
    \caption{Normalized confusion matrices of everyday activities (left) and head motions (right) in leave-one-participant-out user evaluation of \textbf{semi-in-the-wild study}.}
    \label{fig:conf-mat-act-hm-wild}
\end{figure}

We evaluated the efficacy and robustness of the \name{} system through a semi-in-the-wild user study, in which participants performed activities and gestures across three real-world scenarios: walking indoors, sitting in a cafe, and walking along the roadside. Each participant contributed nine sessions of data in these naturalistic settings. Using a leave-one-participant-out evaluation protocol, we treated one participant's sessions as test data while using data from all other participants for training. The mean macro F1-score across all participants was 82.65\%, with accuracy rates of 80.92\% for gestures, 82.13\% for everyday activities, and 84.90\% for head motions. The confusion matrices in Figures~\ref{fig:conf-mat-gest-wild}~and~\ref{fig:conf-mat-act-hm-wild} provide a detailed breakdown of performance across activities.

\begin{figure}[h]
    \centering
    \includegraphics[scale=0.5]{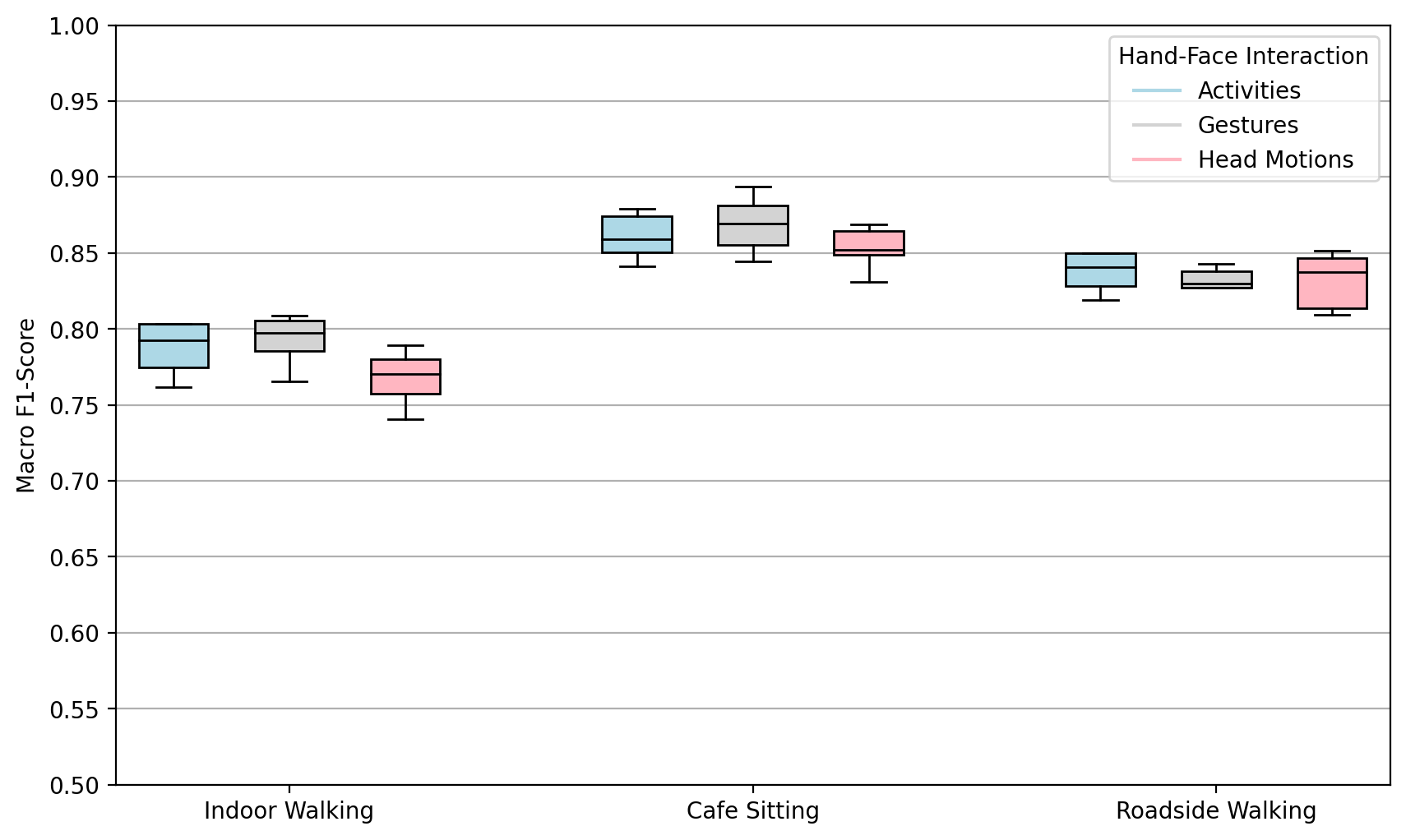}
    \caption{Comparison of performance across various scenarios in the \textbf{semi-in-the-wild user study}.}
    \label{fig:box-plot-wild}
\end{figure}

Compared to the controlled in-lab study, performance in this semi-in-the-wild study was lower for all gestures and activities. This reduction likely results from greater variability in data due to random user movements and the influence of diverse environmental conditions. Figure~\ref{fig:box-plot-wild} sheds light on these factors, showing that walking, whether indoors or outdoors, degrades performance more than sitting in a cafe. However, acoustic noise in the cafe or curbside environments had minimal impact on \name{}’s performance, as most ambient noise occurs in the audible frequency range, whereas \name{} operates in a higher frequency spectrum.

An analysis of sensor readings from this study revealed that walking introduced distinct reflection patterns in the system’s differential echo profile due to the tandem movement of hands. Additionally, when participants walked near objects, these reflections were occasionally captured, potentially affecting recognition accuracy. As a result, subtle gestures and activities in the \name{} recognition set were sometimes overshadowed by reflection patterns associated with walking, leading to occasional errors by the deep learning model. Further analysis indicated that this imputation effect varies across users.

To mitigate these effects, we fine-tuned the user-independent model from the semi-in-the-wild data with one session from the test user (approximately 9.5 minutes of data). This personalization improved the mean macro F1-score across participants from the initial 82.65\% to 88.51\%. These findings suggest that imputation issues from locomotion and environmental factors can be reduced by tailoring the acoustic data model to individual users, enabling more accurate tracking of fine-grained hand-face interactions.

\section{Discussion}
\label{sec:discussion}

\subsection{Performance Comparison of Different Deep Learning Encoders}

\begin{table}[h!]
\begin{tabular}{@{}lccc@{}}
\toprule
\multicolumn{1}{c}{\textbf{Model Architecture}} & \textbf{\begin{tabular}[c]{@{}c@{}}Number of Parameters \\ (Approx.)\end{tabular}} & \textbf{\begin{tabular}[c]{@{}c@{}}In-lab Study \\ Macro F1-score\end{tabular}} & \textbf{\begin{tabular}[c]{@{}c@{}}Semi-in-the-wild \\ Macro F1-score\end{tabular}} \\ \midrule
MobileNetV2~\cite{sandler2018mobilenetv2}                                     & 4M                                                                                 & 82.04\%                                                                         & 74.41\%                                                                             \\
ResNet18~\cite{resnet}                                        & 11M                                                                                & 81.38\%                                                                         & 76.22\%                                                                             \\
2DConv-LSTM                                     & 16M                                                                                & 74.27\%                                                                         & 67.49\%                                                                             \\
\textbf{\name{} Transformer}                            & \textbf{10M}                                                                       & \textbf{93.08\%}                                                                & \textbf{82.65\%}                                                                    \\ \bottomrule
\end{tabular}
\caption{Comparison of \name{} performance under different deep learning frameworks. The number of trainable parameters for each model is reported in millions (M).}
\label{tab:ml-comp}
\end{table}

The deep learning pipeline of \name{} leverages self-attention-based modeling on sliding windows of differential echo profiles to capture active acoustic reflections. Using a transformer encoder~\cite{Transformer, ViT}, the system learns the spatiotemporal patterns of hand movements and associated facial deformations resulting from hand-face interactions. Table~\ref{tab:ml-comp} presents a performance comparison between the \name{} transformer encoder and other convolutional and recurrent network-based approaches. 

While MobileNetV2~\cite{sandler2018mobilenetv2} offers a lightweight architecture with fewer parameters, it underperforms in tracking fine-grained hand-face interactions in both controlled and semi-in-the-wild settings. Meanwhile, ResNet18~\cite{resnet} and a custom 2D convolutional network combined with LSTM layers, despite having more trainable parameters than the transformer encoder, fall short in tracking accuracy. This comparison highlights the transformer encoder's~\cite{Transformer, ViT} unique balance between performance and model efficiency. Furthermore, the transformer architecture’s ability to model longer dependencies and parallel processing, unlike other convolutional or recurrent architectures, makes it particularly well-suited for processing data from the \name{} sensing system.

\subsection{Impact of High-Frequency Components from Activity and Environmental Sounds}
While \name{} operates in the ultrasonic range, which typically excludes sounds generated by human activities and environmental noise, there have been occasional instances, both in lab and naturalistic settings, where the system captures high-frequency components from activities and surroundings. For example, we observed instances where strong coughing or sneezing, especially when the microphone was close to the user’s mouth, introduced high-frequency noise. Additionally, noise was occasionally detected when the sensor module made contact with clothing or hair, resulting in noise artifacts in the reflection pattern due to high-frequency components.

These noisy reflection patterns can lead to incorrect inferences by the deep learning model. To improve robustness, we applied random Gaussian noise as a data augmentation technique in the training set, which enhanced the model's performance. However, achieving consistent robustness against these outliers will require a larger dataset that encompasses a wider variety of noise conditions.

\subsection{Impact of Covering the Device with a Sleeve}

As with other wrist-mounted sensing systems, such as cameras or mmWave radars, the active acoustic sensing system in \name{} faces challenges when covered by the user’s sleeve. Since \name{} relies on acoustic reflections, the presence of a sleeve can introduce noise into the signal. When the sleeve covers the device, it touches both the speaker and microphone, causing distortion in the transmitted ultrasonic wave and introducing high-frequency scratching noise.

To measure the impact of sleeve coverage, we conducted a pilot study with one researcher. In the study, the researcher wore a long-sleeved jacket that covered the \name{} device while performing various activities and gestures. Results showed a significant decrease in macro F1-score in user-dependent evaluations, dropping from 96.4\% to 36.9\%. Analysis of the differential echo profiles—the input to the machine learning pipeline—revealed that sleeve-covered patterns contained arbitrary noise and lacked the consistent motion patterns observed in uncovered data.

One potential solution is to integrate an IMU or ultrawideband (UWB) sensor alongside the \name{} sensing module, providing low-fidelity hand movement information when \name{}'s fine-grained sensing is disrupted by sleeve coverage. However, the feasibility and effectiveness of this approach require further exploration, which we leave for future work.

\subsection{User Comfort and Privacy Preservation}
Unlike passive acoustic sensing methods that capture audible sound from the surroundings, \name{} relies on ultrasonic acoustic waves, which do not inherently contain sensitive information. However, to process the reflection patterns effectively, deep learning inference must be performed on a cloud server with GPU capabilities, meaning that some user data must leave the wearable device. This may raise privacy concerns. Implementing the machine learning model directly on a hardware-accelerated microcontroller could enhance privacy, a solution we aim to explore in future work.

Another consideration is user comfort. While most smartwatches already include a speaker and microphone, implementing this sensing system for optimal hand-face tracking requires positioning the transceivers on the smartwatch strap, which may necessitate slight modifications to off-the-shelf smartwatches. User comfort with the \name{} smartwatch prototype was assessed using a Likert scale from 0 to 5, where 0 indicated “most uncomfortable” and 5 “most comfortable”. Across all 15 study participants, the mean comfort rating was $3.933 \pm 0.799$.

\subsection{Improving Performance in Real-World Scenarios}
We observe a decrease in \name{}'s performance when transitioning from a controlled lab setting to semi-real-world scenarios. This decrease in accuracy is primarily due to motion artifacts in the reflection patterns caused by users walking while performing gestures and activities. A straightforward approach to address this issue would be to expand the training dataset, allowing the model to learn generalized patterns of user movement. However, a multimodal approach may offer a more efficient solution than simply enlarging the dataset.

In this multimodal approach, an IMU—a very low-power sensor—could be used to track user locomotion. With this data, the reflection patterns from gestures and activities could be separated from those caused by walking or other motions. Contrastive learning could further facilitate this disentanglement of patterns. However, this approach would still require a larger dataset with synchronized readings from multiple sensing modalities, including the active acoustic data utilized in \name{}. We leave this solution for future exploration.

\subsection{Potential Applications}
\subsubsection{Tracking Health-Related Behaviors}
\name{} has shown strong potential for tracking a variety of hand-face gestures and everyday activities, including drinking, brushing teeth, and coughing, which are important markers for monitoring health. For example, \name{}'s ability to accurately track drinking behavior in both controlled and naturalistic settings could be extended to estimate a user’s fluid intake based on the duration of drinking events. Similarly, tracking brushing behavior could help users monitor their dental care routines, providing reminders if an event is missed. Additionally, \name{} could support public health by monitoring frequent hand-to-face touches, which is particularly relevant for sanitation concerns in the post-pandemic era.

\subsubsection{Enhancing Interaction in Augmented Reality Environments}
\name{}’s ability to detect fine-grained hand-face gestures and head motions makes it a promising input modality for augmented reality (AR) tasks. It could enable users to perform basic AR actions such as confirming selections, zooming, and switching modes using natural gestures, thus allowing more seamless and intuitive interactions with AR headsets and other devices.

\subsection{Limitations and Future Work}

\subsubsection{Midas Touch Problem}
Humans often touch their faces without any specific intention, making it likely that unintentional gestures included in \name{}'s tracking set may be detected. In such cases, the user does not actually intend to interact with the smartwatch. To address this issue, a trigger motion could be designed to activate \name{}'s hand-face interaction sensing only when intentional interaction is expected. This approach could reduce the likelihood of detecting unintentional gestures, leading to more efficient use of computational and power resources in real-world scenarios.

\subsubsection{Wearing the \name{} Smartwatch on the Non-Dominant Hand}
During user studies, participants wore the \name{} smartwatch on their dominant hand and performed gestures and activities with this same hand. However, many smartwatch users prefer wearing the device on their non-dominant hand to minimize interference with daily activities. In our preliminary evaluation, we found that the acoustic reflection patterns were less distinct for some subtle gestures and activities when \name{} was worn on the non-dominant hand. Achieving similar performance to that of the dominant hand placement may require further evaluation of acoustic signal strength, denoising, and modeling techniques. We plan to explore this aspect in future work.

\section{Conclusion}

This paper introduces \name{}, a wrist-worn active acoustic sensing system designed to capture fine-grained hand-face interactions with minimal intrusion. Through two user studies totaling 15 participants—conducted in both controlled and semi-in-the-wild settings—\name{} demonstrates robust performance, achieving Macro F1-scores of 93.08\% in a lab setting and 82.65\% in semi-in-the-wild environments. We envision \name{} as a straightforward and privacy-aware addition to smartwatches, enabling the seamless capture of hand-face interactions.


\bibliographystyle{ACM-Reference-Format}
\balance
\bibliography{main}

\end{document}